# Introduction to Lattice Gaugefixing
# and Effective Quark and Gluon Masses

Claude Bernard

*Physics Dept., Washington University, St. Louis, MO 63130, USA*

Amarjit Soni and Ken Yee[1]

*Physics Dept., Brookhaven National Laboratory, Upton, NY 11973, USA*

## ABSTRACT

We report on the status of quark and gluon propagators in quenched, gaugefixed lattice QCD. In Landau gauge we find that the effective quark mass in the chiral limit is $M_q \sim 350(40)MeV$. Quark and gluon propagators, the slope of the quark dispersion relation, and effective masses all appear to depend on gauge. A link-chain picture of lattice gaugefixing in the color $N \to \infty$ and strong coupling limit, where the system becomes almost solvable, supports the gauge variance of these numerical results.

## 1. Motivation and Overview

Lattice gaugefixing has drawn considerable attention in recent years. Gluon [1], quark and photon propagators and effective masses have been evaluated on the lattice in Landau, Coulomb and related "λ-gauges," $\lambda \partial_0 A^0 + \nabla \cdot \vec{A} = 0$. Wavefunctions of valence quarks inside mesons have been probed with gauge-variant extended sources. Furthermore, lattice gaugefixing is routinely performed nowadays in order to use extended gauge-variant sources as interpolating fields in the calculation of gauge-*in*variant quantities. While beyond the scope of this talk (see Refs. [2]-[4] for some references), the numerical work has spawned analytical and computational studies of longitudinal and topological gaugefixing ambiguities and their effects on gluon, quark and photon correlation functions and operator product expansion coefficients determined from gauge covariant matching conditions.

For perspective let us focus momentarily on the solvable Schwinger model, where quarks(electrons) are confined but the photon is physical and has a gauge-invariant nonzero mass from the $U_A(1)$ anomaly. Coulomb gauge—the unitary gauge for the Schwinger model—has no unphysical modes and the Coulomb gauge quark propagator is the physical amplitude for quark propagation. Since in $1 + 1$ dimensional empty space a quark's electric field $\vec{E}$ is constant and cannot die away, $\vec{E}$ produces a constant flux of quark-antiquark pairs moving away from the original quark at the speed of light. Consequently the quark propagator—the amplitude to have just one quark at $x \neq 0$ starting with an $x = 0$ quark—vanishes and the effective *Coulomb* gauge quark mass diverges. On the other hand, the *covariant* gauge quark propagator is not the physical amplitude for quark propagation. In fact the Landau gauge quark mass vanishes. This discrepancy with Coulomb gauge is due to unphysical modes (or, alternatively, Gupta-Bleuler physical-state conditions on the Hilbert space) which ruin the identification of the covariant gauge quark propagator with the physical propagation amplitude. Thus, the effective quark mass can be shown to *vary* with covariant gauge parameter [4]. The photon mass, being physical, is gauge *in*variant.

The point is, in association with confinement, the effective quark mass is ambiguous because it varies with gauge—at least in the Schwinger model. Like Schwinger model quarks, QCD quarks and gluons are confined and unphysical and nothing guarantees their masses be gauge-invariant. If what is not forbidden occurs (as usually happens), quark and gluon masses would be gauge-variant.

## 2. Numerical Lattice QCD Results

Numerical quark and gluon propagators at large times—as large as our lattices permit—look approximately, but not exactly, like free particle propagators.[2] In $\lambda$-gauges $M_q$ and $M_g$ are evaluated by matching gaugefixed $\vec{p} = 0$ propagators to free particle propagators at large Euclidean time $t$. In particular, the zero momentum $\lambda$-gauge $M_q^{(\lambda)}$ is obtained by comparing the numerical quark propagator to $Z_q(\frac{1+\gamma^0}{2})\mathrm{e}^{-M_q^{(\lambda)}t}$. The "$(\lambda)$" superscript allows for $\lambda$ dependence. Since the scalar "1" part of the numerical quark propagator equals the vector "$\gamma^0$" part at large enough times, we can focus on the scalar part without loss of generality. Correlated $\chi^2$-fits of the large time scalar part of lattice quark propagators to $Z_q\mathrm{e}^{-M_q^{(\lambda)}t}$ give unacceptably large $\chi^2$s. Quark propagators fit well to trial functions like $At^B\mathrm{e}^{-m_B t}$.

Rather than use an unmotivated but better-fitting trial function, in the rest of this talk we will ignore the large $\chi^2$'s and assume free particle form $Z_q\mathrm{e}^{-M_q^{(\lambda)}t}$. Then the following features are observed: (i)The Dirac scalar component of the $\vec{p} = 0$ numerical quark propagator equals the vector component at large times. (ii)As depicted in Figure 1, quark and gluon effective masses vary with gauge parameter $\lambda$. They decrease with increasing $\lambda$ such that most of the change occurs when $0 < \lambda \leq 2$. As $\lambda \to \infty$, the masses stabilize to a nonzero value. (iii)Motivated by the chiral perturbation theory relation $M_\pi^2 \propto m_q$, a fit to

$$M_q^{(\lambda)} = b^{(\lambda)}M_\pi^2 + M_c^{(\lambda)} \tag{1}$$

in the chiral regime yields in Landau gauge at $\beta = 6.0$ that $b^{(1)} \sim 2.7(.3) \times 10^{-4}/\mathrm{MeV}$ and $M_c^{(1)} \sim 350(40)\mathrm{MeV}$. In general, $M_q$ is about one-half the $\rho$ mass at the same hopping constant $K$, the lattice bare mass parameter, independently of $K$. (iv)$M_g$ is about 70% larger than $M_c^{(1)}$. (v)The slope of the dependence of quark energy square, $E_q^2$, on spatial momentum square, $\vec{p}^2$, varies with $\lambda$ so that not all gauges can be consistent with Lorentz invariance.

---

[2] The numerical calculation is on $\beta = 5.7$ and 6.0 quenched Wilson lattices. For noncogniscenti, $\beta$ is a measure of physical lattice spacing; the larger $\beta$ is, the smaller the lattice spacing and the closer one is to the continuum limit. $\beta = 6.0$ correspond to a lattice spacing of $\sim 1.7 GeV^{-1}$.

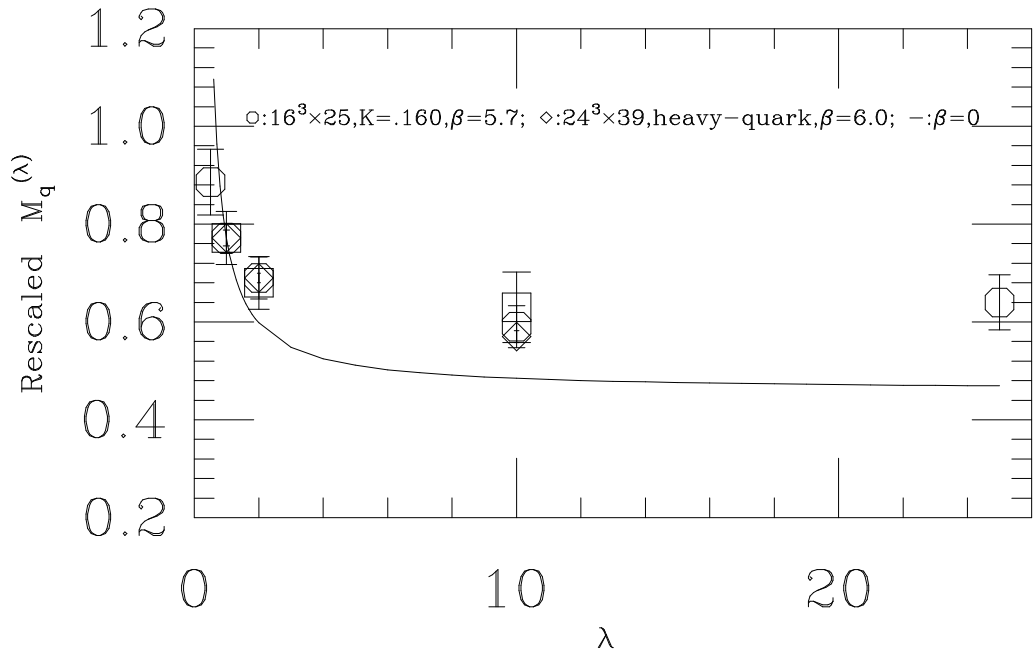

FIG. 1. Comparison of $M_q^{(\lambda)}$ to Strong Coupling $\lambda$-Dependence

## 3. Lattice Gaugefixing and $\beta = 0$, $N \to \infty$ Model

The $\lambda$ dependences of the effective quark mass and dispersion relation are qualitatively reproduced in a $\beta = 0$, color $N \to \infty$ analytical solution of $\lambda$-gauge lattice QCD. Figure 1 compares the chiral quark mass $M_c^{(\lambda)}$ at strong coupling to the numerical effective quark masses rescaled so that they all agree at $\lambda = 1$. The strong coupling calculation was done in the following way. $\lambda$-gauge vacuum expectation value $\langle \Theta \rangle$ in lattice QCD is defined as

$$\langle \Theta \rangle \;\equiv\; \Big[\,\big[\,[\Theta]_f\big]_v\,\Big]_u, \quad [\Theta]_\theta \;\equiv\; z_\theta^{-1} \int [d\theta] \mathrm{e}^{-S^\theta} \Theta, \quad z_\theta \equiv \int [d\theta'] \mathrm{e}^{-S^{\theta'}}, \tag{2}$$

where labels $\theta = f$, $v$ or $u$ refer respectively to quark, gauge transformation, and lattice gauge "link" fields. $[\Theta]_\theta$ is the average of the operator (or combination of operators) $\Theta$ over the fields $\theta = f$, $v$, or $u$. $f$ stands for quark fields $\psi$; $v$, for fields $V$ which transform a background gauge field into $\lambda$-gauge; and $u$, for the lattice gauge fields $U$. $S^f$ and $S^u$ are the usual lattice Wilson fermion and plaquette actions. The lattice gaugefixing action $S^v$ is

$$S^v \;\equiv\; -\sum_{x,\mu} \frac{J_n}{2}\, \mathrm{Retr}(V_x U_{x,\mu} V_{x+\hat\mu}^\dagger), \qquad J_\mu \equiv \frac{\lambda_\mu}{\xi} \tag{3}$$

where $\lambda_0 \equiv \lambda$ and $\lambda_i \equiv 1$. Quarks are quenched by choice; consistency with the Fadeev-Popov method requires $\{V_x\}$ to be quenched. $S^v$ is minimized site-by-site with respect to $\{V_x\}$ in numerical simulations to achieve $\lambda$-gauge. Similarly, the $\xi \to 0$ limit corresponds to $\lambda$-gauge.

Since this system is not solvable, we take the $\lambda$-gauge, $N \to \infty$ limit [3]. Identifying $\xi \equiv 1/(2N)$, taking $N \to \infty$ and integrating out gaugefixing fields $\{V_x\}$ yields

$$\langle V_y(\,_y\mathcal{L}_w)V_w^\dagger\rangle_{\alpha j} = \delta_{\alpha j}\sum_{w\tilde{\mathcal{L}}_y} A_{w\tilde{\mathcal{L}}_y}(J)\ \mathrm{tr}[\,_y\mathcal{L}_w\ _w\tilde{\mathcal{L}}_y]_u, \quad A_{w\tilde{\mathcal{L}}_y} = \frac{1}{N^2}\mathrm{tr}\langle V_{yy}\tilde{\mathcal{L}}_w V_w^\dagger\rangle_{|\beta=0}. \quad (4)$$

$\alpha$ and $j$ are color indices ranging from 1 to $N$ and $_y\mathcal{L}_w$ is a link chain from $w$ to $y$. Eq. (4) says the gaugefixed expectation value of $_y\mathcal{L}_w$ is the weighted sum over all Wilson loops made by joining to $_y\mathcal{L}_w$ a selfavoiding link segment $_w\tilde{\mathcal{L}}_y$. (4) identifies weights $A_{w\tilde{\mathcal{L}}_y}$ with the $\beta = 0$ gaugefixed expectation value of $_w\tilde{\mathcal{L}}_y$. All of the $\lambda$ dependence is in these weights.

Hopping expanding the quark fields, applying (4), and resumming the hopping expansion leads to an expression for the quark propagator. In $D = 3 + 1$ dimensions in the color $N \to \infty$ and $\beta = 0$ limit, and in an orthogonal trace approximation described in Ref. [3], the quark propagator dispersion relation is

$$E_q^{(\lambda)2} = M_q^{(\lambda)2} + \sum_i \overline{g}_i^2 p_i^2, \qquad \overline{g}_i(\lambda) = \begin{cases} \frac{3}{4\lambda} & \lambda \leq \frac{1}{2}; \\ \frac{4\lambda-1} & \lambda \geq \frac{1}{2}. \end{cases} \quad (5)$$

Since slope $\overline{g}_i(\lambda \neq 1) \neq 1$, the dispersion relation is not free-particle form except in Landau gauge. At $\vec{p} = 0$ $\overline{g}_i$ dependence drops out, but $M_q^{(\lambda)}$ remains $\lambda$-dependent. The quark propagator in this limit (with Wilson parameter $r = 1$) has the free particle exponential form.

While we do not explain Wilson parameter $r$ here, let us note that in strong coupling the $r = 0$ point (where backtracking and hence spontaneous chiral symmetry breaking is allowed) is more similar than $r = 1$ to continuum QCD. At $r = 0$ $M_q^{(\lambda)}$ varies with $\lambda$, diverges in Coulomb gauge, and is analytic and linear in $m_q$ near $m_q \to 0$. Following (1), the Taylor expansion $M_q^{(\lambda)} = M_c^{(\lambda)} + B^{(\lambda)}m_q + \mathcal{O}(m_q^2)$ yields

$$M_c^{(\lambda)} = \frac{\sqrt{7}}{242\lambda}\begin{cases} 67 - 25\lambda^2 & \lambda \leq \frac{1}{2}; \\ \frac{-25+200\lambda+672\lambda^2}{4(4\lambda-1)} & \lambda \geq \frac{1}{2}; \end{cases} \quad B^{(\lambda)} = \begin{cases} \frac{20203+4753\lambda^2}{29282\lambda} & \lambda \leq \frac{1}{2}; \\ \frac{4753-38024\lambda+399296\lambda^2}{117128\lambda(4\lambda-1)} & \lambda \geq \frac{1}{2}. \end{cases} \quad (6)$$

## 4. Acknowledgements

Computing was at NERSC (partially within "Grand Challenge") and SDSC. AS and KY are partially supported by DOE contract $DE - AC02 - 76CH00016$; CB partially by DOE grant $DE2 - FG02 - 91ER40628$.